\documentclass[prd,aps,twocolumn,showpacs,nofootinbib,amsmath,floatfix]{revtex4}

\usepackage{times}
\usepackage{hyperref}

\usepackage{graphicx}
\usepackage{dcolumn}
\usepackage{bm}

\begin{document}

\title{Radiative corrections to the Dalitz plot of $K_{l3}^\pm$ decays: Contribution of the four-body region}

\author{J.\ J.\ Torres}

\affiliation{Departamento de Formaci\'on B\'asica, Escuela Superior de C\'omputo del IPN, Apartado Postal 75-702, M\'exico, D.F.\ 07738, Mexico}

\author{A.\ Mart{\'\i}nez}

\affiliation{Departamento de F{\'\i}sica, Escuela Superior de F\'{\i}sica y Matem\'aticas del IPN, Apartado Postal 75-702, M\'exico, D.F.\ 07738, Mexico}

\author{M.\ Neri}

\affiliation{Departamento de F{\'\i}sica, Escuela Superior de F\'{\i}sica y Matem\'aticas del IPN, Apartado Postal 75-702, M\'exico, D.F.\ 07738, Mexico}

\author{C.\ Ju\'arez-Le\'on}

\affiliation{Departamento de F{\'\i}sica, Escuela Superior de F\'{\i}sica y Matem\'aticas del IPN, Apartado Postal 75-702, M\'exico, D.F.\ 07738, Mexico}

\author{Rub\'en Flores-Mendieta}

\affiliation{Instituto de F{\'\i}sica, Universidad Aut\'onoma de San Luis Potos{\'\i}, \'Alvaro Obreg\'on 64, Zona Centro, San Luis Potos{\'\i}, S.L.P.\ 78000, Mexico}

\date{\today}

\begin{abstract}

We calculate the radiative corrections to the Dalitz plot of $K_{l3}^\pm$ decays to order $(\alpha/\pi)(q/M_1)$, where $q$ is the momentum transfer and $M_1$ is the mass of the kaon. We restrict the analysis to the so-called four-body region, which arises when no discrimination of real photons is made either kinematically or experimentally. We present our results in two ways: the first one with the triple integration over the photon kinematical variables ready to be performed numerically and the second one in a fully analytical form. Our results can be useful in experimental analyses of the Dalitz plot, by evaluating the model-independent coefficients of the quadratic products of the form factors; we provide some numbers as a case example. We find a small, albeit non-negligible, contribution from the four-body region to the radiative correction to the total decay rate of $K_{l3}^\pm$ decays.

\end{abstract}


\maketitle

There are several inherent difficulties in the analysis of radiative corrections (RC) in $K_{l3}$ decays. The absence of first principles to evaluate them introduces a model dependence so experimental analyses which use them also become model-dependent. RC depend on the process characteristics, such as the charge assignment of the participating mesons, the size of the momentum transfer, and whether real photons can be experimentally discriminated or not. RC also depend on the observable to be measured. All this requires RC to be recalculated everytime the process characteristics and/or the observable are changed. Finally, there are difficulties of a practical nature: It turns out that the final expressions of RC calculations are rather inefficient to use or are long and tedious to the point that their use becomes unreliable.

In a recent paper \cite{juarez11} we overcome the above difficulties rather satisfactorily. We obtained a model-independent expression for the Dalitz plot of $K_{l3}^\pm$ decays including RC of order $(\alpha/\pi)(q/M_1)$, where $q$ is the momentum transfer and $M_1$ is the mass of the kaon. We studied the so-called three-body region (TBR) of the Dalitz plot. We thus assumed that not only real photons are not detected but also that events whose energies do not satisfy the three-body energy-momentum conservation restrictions are rejected.

When real photons cannot be discriminated, however, the TBR of the Dalitz plot should be extended to the four-body region (FBR). The purpose of the present paper is to extend, on the same footing, the calculations of Ref.~\cite{juarez11} to evaluate the four-body contribution of the RC to the Dalitz plot. Although this problem keeps a close similarity with the one discussed in Ref.~\cite{juarez11}, it is not possible to find a single rule which allows us to adapt one calculation into the other. Thus, we need to take a few steps backwards and start the calculation at the bremsstrahlung transition amplitude level, namely, by squaring such amplitude and then performing the integrals over the kinematical variables of the photon restricted to the FBR. In order to save a substantial amount of effort we will follow the approach implemented in the FBR analysis of RC in baryon semileptonic decays \cite{mar01,rfm02,torr06}.

For definiteness, let us consider the four-body decay
\begin{equation}
K^+(p_1) \to \pi^0(p_2) + \ell^+(l) + \nu_\ell(p_\nu) + \gamma(k), \label{eq:decay}
\end{equation}
where $K^+$ and $\pi^0$ denote a positively charged kaon and a neutral pion, respectively, $\ell^+$ and $\nu_\ell$ denote a positively charged lepton ($\ell^+ = e^+$ or $\mu^+$) and its neutrino, respectively, and $\gamma$ represents a real photon. Here $p_1=(E_1,{\mathbf p}_1)$, $p_2=(E_2,{\mathbf p}_2)$, $l=(E,{\mathbf l})$, $p_\nu=(E_\nu,{\mathbf p}_\nu)$, and $k=(\omega,\mathbf{k})$ are the four-momenta of $K^+$, $\pi^0$, $\ell^+$, $\nu$, and $\gamma$, respectively. $M_1$, $M_2$, and $m$ are the nonzero masses of the first three particles. In the center-of-mass frame of $K^+$, $M_1 = E+E_2+E_\nu+\omega$ and $\mathbf{0} = \mathbf{p}_2 + \mathbf{l} + \mathbf{p}_\nu + \mathbf{k}$, so $E_\nu = E_\nu^0-\omega$ and $\mathbf{p}_\nu = \mathbf{p}_\nu^0 - \mathbf{k}$, where $E_\nu^0$ is the energy  and $\mathbf{p}_\nu^0$ is the three-momentum of the neutrino in the nonradiative process. In addition, $p_2$, $l$, $p_\nu$, and $k$ will also denote the magnitudes of the corresponding three-momenta when the expressions involved are not manifestly covariant. All other conventions and notation are given in Ref.~\cite{juarez11}.

The calculation of bremsstrahlung RC in the FBR simplifies because the events in this region have the same amplitude $\mathsf{M}_B$ as in the TBR, given by Eq.~(41) of Ref.~\cite{juarez11}. The bremsstrahlung differential decay rate in the FBR is then
\begin{eqnarray}
d\Gamma_B^{\mathrm{FBR}} & = & \frac{1}{(2\pi)^8} \frac{1}{2M_1}\frac{mm_\nu}{4E_2EE_\nu\omega} d^3p_2\,d^3l\,d^3p_\nu\,d^3k \, \nonumber \\
&  & \mbox{} \times \delta^4(p_1-p_2-l-p_\nu-k) \sum_{\textrm{spins},\epsilon}|\mathsf{M}_B|^2. \label{eq:diffdgb}
\end{eqnarray}
In order to perform the integrals over the kinematical variables in Eq.~(\ref{eq:diffdgb}), we orient the coordinate axes in such a way that $\ell^+$ is emitted along the $+z$ axis and $\pi^0$ is emitted in the first or fourth quadrant of the $(x,z)$ plane. Therefore, there are five relevant variables of the final state. Two of them are the energies $E$ and $E_2$, whose allowed values in the FBR are given by Eq.~(5) of Ref.~\cite{juarez11}. The other three variables can be grouped into
either $(k,\cos\theta_k,\varphi_k)$ or $(\cos\theta_2,\cos\theta_k,\varphi_k)$, where $k$, $\theta_k$, and $\varphi_k$ are the magnitude of the three-momentum and the polar and azimuthal angles of the photon, respectively, and $\theta_2$ is the polar angle of $\pi^0$. If we define $\hat{\mathbf{p}}_2\cdot\hat{\mathbf{l}} = \cos\theta_2 \equiv y$ and $\hat{\mathbf{l}} \cdot \hat{\mathbf{k}} = \cos\theta_k \equiv x$, the photon energy is given by \cite{juarez11}
\begin{equation}
\omega = \frac{F}{2D}, \label{eq:defw}
\end{equation}
where
\begin{subequations}
\begin{eqnarray}
F & = & 2p_2l(y_0-y), \\
D & = & E_\nu^0 + lx+\mathbf{p}_2\cdot\hat{\mathbf{k}},
\end{eqnarray}
\end{subequations}
with
\begin{equation}
y_0 = \frac{{E_\nu^0}^2-p_2^2-l^2}{2p_2l}. \label{eq:y0}
\end{equation}

By following a standard procedure, the differential decay rate in the FBR can be written as
\begin{eqnarray}
d\Gamma_B^{\mathrm{FBR}} & = & \frac{\alpha}{\pi} d\Omega \Bigg\{ A_0 I_{0F}(E,E_2) \nonumber \\
&  & \mbox{} + \frac{p_2l}{4\pi}\int_{-1}^1 dx \int_{-1}^1 dy \int_0^{2\pi} d\varphi_k |\sf{M}_B^\prime|^2 \Bigg\}. \label{eq:dfbr}
\end{eqnarray}

The organization of Eq.~(\ref{eq:dfbr}) allows one to contrast it with its counterpart in the TBR, constituted by Eqs.~(68)-(73) of Ref.~\cite{juarez11}, where the factors $d\Omega$ and $A_0$ are defined. The counterpart
of the first summand on the right-hand side of Eq.~(\ref{eq:dfbr}) is the product $A_0I_0$, where the function $I_0$, defined in Eq.~(65) of this reference, contains the infrared divergence. Now, $I_0$ is replaced by the infrared-convergent function $I_{0F}$ \cite{mar01,rfm02,torr06} defined as
\begin{equation}
I_{0F} = \frac{\theta_{0F}}{2}\ln \left(\frac{y_0+1}{y_0-1} \right), \label{eq:iof}
\end{equation}
where
\begin{equation}
\theta_{0F} = 4 \left(\frac{1}{\beta} \mathrm{arctanh}\, \beta-1 \right),
\end{equation}
and $\beta = l/E$. All other contributions of $|{\sf M}_B|^2$ can be cast into a single term $|{\sf M}_B^\prime|^2$, whose form is not needed here. Let us notice that the variable $y$ falls within the intervals $[-1,y_0]$ and $[-1,1]$ in the TBR and FBR, respectively.

Performing the integrals in Eq.~(\ref{eq:dfbr}) yields
\begin{widetext}
\begin{eqnarray}
d\Gamma_B^{\mathrm{FBR}} & = & \frac{\alpha}{\pi} d\Omega \Bigg\{ A_0 I_{0F}(E,E_2) + \sum_{i=1}^5 \frac{8}{M_1^2} \Big[ C_i^+ |f_+|^2 + C_i \mathrm{Re} [f_+ f_-^*] + C_i^-|f_-|^2 \Big] \Bigg\}, \label{eq:dBfinal}
\end{eqnarray}
\end{widetext}
where, for simplicity, hereafter we will use the shorthand notation $f_\pm \equiv f_\pm(q^2)$ unless explicitly noted otherwise. Let us stress that there is a one-to-one correspondence between the five terms involved in the second expression on the right-hand side of Eq.~(\ref{eq:dBfinal}) and Eqs.~(69)-(73) of Ref.~\cite{juarez11}, so we have kept the parallelism between both calculations all along. The coefficients $C_i$ read
\begin{eqnarray}
\begin{array}{ll}
C_1^+ = \Lambda_{1F} + \Lambda_{2F} + \Lambda_{3F},
&
C_1 = -\Lambda_{2F} - 2\Lambda_{3F},
\\
C_1^- = \Lambda_{3F},
&
C_2^+ = \Lambda_{4F} + \Lambda_{5F} + \Lambda_{6F},
\\
C_2 = - \Lambda_{5F} - 2\Lambda_{6F},
&
C_2^- = \Lambda_{6F},
\\
C_3^+ = \Lambda_{7F} + \Lambda_{8F},
&
C_3 = \Lambda_{7F},
\\
C_3^- = -\Lambda_{8F},
&
C_4^+ = \Lambda_{9F} + \Lambda_{10F},
\\
C_4 = \Lambda_{9F},
&
C_4^- = -\Lambda_{10F},
\\
C_5^+ = C_5^- = C_5/2 = \Lambda_{11F}.
&
\end{array}
\end{eqnarray}

The functions $\Lambda_{kF}$ ($k=1,\ldots,11$) are the result of performing the triple integrals contained in Eq.~(\ref{eq:dfbr}). They have the very same structure as their TBR counterparts listed in Appendix B of Ref.~\cite{juarez11}, except for the fact the upper limit of integration over the variable $y$ is replaced by 1.

Equation (\ref{eq:dBfinal}) constitutes our first result: an expression for the bremsstrahlung RC to the $K_{l3}^\pm$ Dalitz plot to order $(\alpha/\pi)(q/M_1)$, restricted to the FBR. At this stage of the calculation, $d\Gamma_B^{\textrm{FBR}}$ contains triple integrals over the kinematical variables of the photon, which can be performed numerically.

We can proceed further, however, and provide a fully integrated expression that could be more useful in a Monte Carlo simulation. For this task, we use previous results obtained in the analysis of the FBR of the Dalitz plot of baryon semileptonic decays \cite{mar01,rfm02,torr06}. Accordingly, the analytical form of Eq.~(\ref{eq:dBfinal}) can be expressed as
\begin{equation}
d\Gamma_B^{\mathrm{FBR}} = \frac{\alpha}{\pi} d\Omega \left[ A_{1F} |f_+|^2 + A_{2F} \mathrm{Re}(f_+f_-^*) + A_{3F} |f_-|^2 \right], \label{eq:dG3}
\end{equation}
where
\begin{equation}
A_{iF} = A_{i}^{(0)}I_{0F} + A_{iF}^{(B)}, \qquad i=1,2,3. \label{eq:Aifprima}
\end{equation}
The functions $A_{i}^{(0)}$ are defined in Eqs.~(17)-(19) of Ref.~\cite{juarez11}, whereas the functions $A_{iF}^{(B)}$ read
\begin{equation}
A_{1F}^{(B)} = \frac{8}{M_1^2}\sum_{i=1}^{11} \Lambda_{iF},
\end{equation}
\begin{eqnarray}
A_{2F}^{(B)} & = & \frac{8}{M_1^2} ( -\Lambda_{2F} - 2\Lambda_{3F} - \Lambda_{5F} - 2\Lambda_{6F} + \Lambda_{7F} + \Lambda_{9F} \nonumber \\
&  & \mbox{} + 2\Lambda_{11F}),
\end{eqnarray}
and
\begin{equation}
A_{3F}^{(B)} = \frac{8}{M_1^2} ( \Lambda_{3F} + \Lambda_{6F} - \Lambda_{8F} - \Lambda_{10F} + \Lambda_{11F}).
\end{equation}
where the functions $\Lambda_{kF}$ ($k=1,\ldots,11$) correspond to the analytical versions of their counterparts $\Lambda_k$ of the TBR, Eqs.~(74)-(84) of Ref.~\cite{juarez11}. After some rearrangements, one finds in a close analogy with Eqs.~(87)-(89) of this reference
\begin{widetext}
\begin{eqnarray}
\frac{M_1^2}{4p_2l} A_{1F}^{(B)} & = & \Bigg[ \frac{2E_2}{M_1} + \frac{5E}{M_1} - \frac{E_2E}{M_1^2} - \frac{E^2}{M_1^2}(1+\beta^2) \Bigg] \eta_{0F} + \Bigg[ 1 - \frac{2E}{M_1} + \frac34 \frac{m^2}{M_1^2} \Bigg] \theta_{0F} - \frac{m^2}{E^2} \Bigg[ 2E_\nu^0 + \frac{m^2}{M_1} \Bigg] \theta_{2F} \nonumber \\
&  & \mbox{} +\Bigg[ E_\nu^0-3E-p_2\beta y_0+\frac{m^2}{E} \Bigg[3-\frac{2E_2}{M_1} + \frac{m^2}{M_1^2} - \frac34 \frac{E(E+E_\nu^0)+p_2ly_0}{M_1^2} \Bigg] \Bigg] \theta_{3F} + \Bigg[ E_\nu^0+\frac{2EE_2}{M_1}+\frac{4E^2}{M_1} + \frac{2p_2ly_0}{M_1} \nonumber \\
&  & \mbox{} - \frac{m^2}{4M_1}\Bigg[ 9+\frac{3E_2}{M_1}+\frac{4E}{M_1}\Bigg] \Bigg] \theta_{4F} + \Bigg[ 4+\frac{2E}{M_1}+\frac{2p_2ly_0}{M_1^2}-\frac{m^2}{4M_1^2}\Bigg] l\theta_{5F}+\frac{m^2}{E^2} \theta_{6F} + \Bigg[ -2+
\frac{E_\nu^0}{E} - \frac{m^2}{M_1E}\Bigg] \theta_{7F} \nonumber \\
&  & \mbox{} - \frac{1}{2E} \theta_{9F} + \frac{4l^2E_\nu^0}{M_1^2}\theta_{10F}+\frac{2p_2l}{M_1}
\theta_{12F}-\frac{2p_2l}{M_1} \theta_{13F} + \frac{l(E+E_2)}{M_1^2} \theta_{14F} - \frac{2p_2l^2}{M_1^2} \theta_{19F} + \frac{2l^3}{M_1^2} \theta_{20F}, \label{eq:a1f}
\end{eqnarray}
\begin{equation}
\frac{M_1^{3}}{m^2} \frac{1}{4p_2l} A_{2F}^{(B)} = -\frac{1}{2M_1} \theta_{0F} + \frac{m^2}{E^2} \theta_{2F} - \Bigg[\frac32 + \frac{m^2}{M_1E} + \frac{E_2-p_2\beta y_0}{2M_1} \Bigg] \theta_{3F} + \Bigg[ \frac12 + \frac{2E+E_2}{2M_1} \Bigg] \theta_{4F} + \frac{l}{2M_1} \theta_{5F}, \label{eq:a2f}
\end{equation}
and
\begin{equation}
A_{3F}^{(B)} = \frac{p_2lm^2}{M_1^4} \left[ -\theta_{0F} + (M_1-E_2+\beta p_2y_0) \theta_{3F} - (M_1-E_2) \theta_{4F} - l\theta_{5F} \right]. \label{eq:a3f}
\end{equation}
\end{widetext}

Equation (\ref{eq:dG3}) constitutes our second result. It is an analytical expression for the Dalitz plot of $K_{l3}^\pm$ decays, restricted to the FBR region. It includes RC to order $(\alpha/\pi)(q/M_1)$ and is model independent. The complete RC to the Dalitz plot of $K_{l3}^\pm$ decays without the restriction of eliminating real photons are given by adding $d\Gamma_B^{\textrm{FBR}}$ to $d\Gamma_B^{\textrm{TBR}}$ given by Eq.~(93) of Ref.~\cite{juarez11}.

An application of our formulas in a Monte Carlo analysis requires the numerical evaluation of the functions $A_i$ and $A_{iF}$ all over the Dalitz plot. This can be done by tracing a lattice in the allowed kinematical region and constructing arrays with these functions evaluated at given points $(E,E_2)$. The arrays should feed the Monte Carlo simulation as a matrix multiplication, with the form factors and the energies $(E,E_2)$ varied in each step of the simulation. In Tables \ref{t:a1ke3} and \ref{t:a1km3} we display the numerical values of $(\alpha/\pi)A_1$ and $(\alpha/\pi)A_{1F}$, the latter in boldface characters, for $K_{e3}^+$ and $K_{\mu 3}^+$ decays, respectively, as a case example.

\begingroup
\begin{table*}
\caption{\label{t:a1ke3}Radiative corrections $(\alpha/\pi)A_1 \times 10$ and $(\alpha/\pi)A_{1F} \times 10$ for $K^+ \to \pi^0 + e^+ + \nu_e$ decay. The entries corresponding to the latter are marked in boldface characters. $E$ and $E_2$ are given in $\textrm{GeV}$.}
\begin{ruledtabular}
\begin{tabular}{crrrrrrrrr}
$E_2\backslash E$ & $ 0.0123$ & $ 0.0370$ & $ 0.0617$ & $ 0.0864$ & $ 0.1111$ & $ 0.1358$ & $ 0.1604$ & $ 0.1851$ & $ 0.2098$ \\
\hline
$0.2592$ & $ 0.1533$ & $ 0.1880$ & $ 0.1462$ & $ 0.0668$ & $-0.0286$ & $-0.1220$ & $-0.1949$ & $-0.2246$ & $-0.1726$ \\
$0.2468$ & $\mathbf{0.1028}$ & $ 0.1810$ & $ 0.1580$ & $ 0.0902$ & $ 0.0011$ & $-0.0905$ & $-0.1654$ & $-0.1998$ & $-0.1537$ \\
$0.2345$ & $\mathbf{0.0701}$ & $ 0.1578$ & $ 0.1522$ & $ 0.0962$ & $ 0.0150$ & $-0.0718$ & $-0.1445$ & $-0.1792$ & $-0.1354$ \\
$0.2222$ & $\mathbf{0.0490}$ & $\mathbf{0.0948}$ & $ 0.1429$ & $ 0.0989$ & $ 0.0261$ & $-0.0551$ & $-0.1249$ & $-0.1590$ & $-0.1168$ \\
$0.2098$ & $\mathbf{0.0339}$ & $\mathbf{0.0599}$ & $ 0.1321$ & $ 0.1004$ & $ 0.0363$ & $-0.0391$ & $-0.1056$ & $-0.1389$ & $-0.0977$ \\
$0.1975$ & $\mathbf{0.0228}$ & $\mathbf{0.0383}$ & $\mathbf{0.0662}$ & $ 0.1014$ & $ 0.0461$ & $-0.0233$ & $-0.0863$ & $-0.1186$ & $-0.0779$ \\
$0.1851$ & $\mathbf{0.0145}$ & $\mathbf{0.0236}$ & $\mathbf{0.0375}$ & $\mathbf{0.0743}$ & $ 0.0558$ & $-0.0075$ & $-0.0670$ & $-0.0979$ & $-0.0568$ \\
$0.1728$ & $\mathbf{0.0084}$ & $\mathbf{0.0134}$ & $\mathbf{0.0203}$ & $\mathbf{0.0343}$ & $ 0.0654$ & $ 0.0083$ & $-0.0474$ & $-0.0769$ & $-0.0333$ \\
$0.1604$ & $\mathbf{0.0042}$ & $\mathbf{0.0065}$ & $\mathbf{0.0095}$ & $\mathbf{0.0149}$ & $\mathbf{0.0281}$ & $ 0.0243$ & $-0.0275$ & $-0.0550$ & $-0.0019$ \\
$0.1481$ & $\mathbf{0.0014}$ & $\mathbf{0.0021}$ & $\mathbf{0.0030}$ & $\mathbf{0.0045}$ & $\mathbf{0.0076}$ & $\mathbf{0.0170}$ & $-0.0070$ & $-0.0316$ &         \\
\end{tabular}
\end{ruledtabular}
\end{table*}
\endgroup

\begingroup
\begin{table*}
\caption{\label{t:a1km3}
Radiative corrections $(\alpha/\pi)A_1 \times 100$ and $(\alpha/\pi)A_{1F} \times 100$ for $K^+ \to \pi^0 + \mu^+ + \nu_\mu$ decay. The entries corresponding to the latter are marked in boldface characters. $E$ and $E_2$ are given in $\textrm{GeV}$.}
\begin{ruledtabular}
\begin{tabular}{crrrrrrrrr}
$E_2\backslash E$ & $ 0.1131$ & $ 0.1280$ & $ 0.1429$ & $ 0.1578$ & $ 0.1727$ & $ 0.1876$ & $ 0.2025$ & $ 0.2174$ & $ 0.2322$ \\
\hline
$ 0.2480$ &           &           &           &           & $-0.3664$ & $-0.2633$ & $-0.2285$ & $-0.1879$ & $-0.1003$ \\
$ 0.2361$ &           & $-0.0168$ & $-0.0295$ & $-0.0624$ & $-0.1026$ & $-0.1406$ & $-0.1649$ & $-0.1584$ & $-0.0861$ \\
$ 0.2242$ & $ 0.0397$ & $ 0.0356$ & $ 0.0096$ & $-0.0301$ & $-0.0760$ & $-0.1188$ & $-0.1472$ & $-0.1437$ & $-0.0725$ \\
$ 0.2123$ & $ 0.0345$ & $ 0.0355$ & $ 0.0139$ & $-0.0224$ & $-0.0657$ & $-0.1068$ & $-0.1341$ & $-0.1303$ & $-0.0582$ \\
$ 0.2004$ & $ 0.0148$ & $ 0.0237$ & $ 0.0091$ & $-0.0214$ & $-0.0600$ & $-0.0975$ & $-0.1222$ & $-0.1170$ & $-0.0438$ \\
$ 0.1885$ & $ \mathbf{0.0238}$ & $ 0.0072$ & $ 0.0006$ & $-0.0231$ & $-0.0560$ & $-0.0890$ & $-0.1106$ & $-0.1036$ & \\
$ 0.1766$ & $ \mathbf{0.0127}$ & $ \mathbf{0.0651}$ & $-0.0096$ & $-0.0260$ & $-0.0527$ & $-0.0809$ & $-0.0991$ & $-0.0899$ & \\
$ 0.1647$ & $ \mathbf{0.0067}$ & $ \mathbf{0.0213}$ & $-0.0207$ & $-0.0294$ & $-0.0498$ & $-0.0729$ & $-0.0874$ & $-0.0761$ & \\
$ 0.1528$ & $ \mathbf{0.0030}$ & $ \mathbf{0.0087}$ & $ \mathbf{0.0213}$ & $-0.0332$ & $-0.0470$ & $-0.0649$ & $-0.0758$ & $-0.0652$ & \\
$ 0.1409$ & $ \mathbf{0.0008}$ & $ \mathbf{0.0024}$ & $ \mathbf{0.0052}$ & $ \mathbf{0.0117}$ & $ \mathbf{0.0498}$ & $-0.0574$ & $-0.0663$ &           &  \\
\end{tabular}
\end{ruledtabular}
\end{table*}
\endgroup

As a further application, we can evaluate the total decay rate of $K_{e3}^\pm$ including RC from both the TBR and FBR. It can be written as
\begin{equation}
\Gamma(K_{e3}^\pm) \sim C_K^2 \frac{G_F^2|V_{us}|^2}{128\pi^3}M_1^5 |f_+^{K^+\pi^0}(0)|^2 [I(\tilde{\lambda}_+)+I_F(\tilde{\lambda}_+)], \label{eq:gg}
\end{equation}
where $I(\tilde{\lambda}_+)$ and $I_F(\tilde{\lambda}_+)$ involve the double integration over the energies $E$ and $E_2$ for the TBR and FBR, respectively. Their expressions read
\begin{eqnarray}
I(\tilde{\lambda}_+) & = & \frac{4}{M_1^2} \int_m^{E_m}dE\int_{E_2^{\mathrm{min}}}^{E_2^{\mathrm{max}}}dE_2 \left[ A_1^{(0)} + \frac{\alpha}{\pi} A_1\right] \nonumber \\
&  & \mbox{} \times \left[1+\frac{q^2}{M_{\pi^\pm}^2} \tilde{\lambda}_+ \right]^2 \nonumber \\
& = & h_0 + h_1 \tilde{\lambda}_+ + h_2 \tilde{\lambda}_+^2, \label{eq:ilambda}
\end{eqnarray}
and
\begin{eqnarray}
I_F(\tilde{\lambda}_+) & = & \frac{4}{M_1^2} \int_m^{E_c}dE\int_{M_2}^{E_2^{\mathrm{min}}}dE_2 \left[ \frac{\alpha}{\pi} A_{1F} \right] \nonumber \\
&  & \mbox{} \times \left[1+\frac{q^2}{M_{\pi^\pm}^2} \tilde{\lambda}_+ \right]^2 \nonumber \\
& = & h_{0F} + h_{1F} \tilde{\lambda}_+ + h_{2F} \tilde{\lambda}_+^2, \label{eq:ilambdaF}
\end{eqnarray}
where $f_+^{K^+\pi^0}(q^2)$ has been expanded linearly in $q^2$, with the slope parameter $\tilde{\lambda}_+=0.0328\pm 0.0033$ \cite{cir}.

By using more refined numerical integration routines than in Ref.~\cite{juarez11}, we find that the uncorrected coefficients are $h_0^{(0)}=0.0965$, $h_1^{(0)}=0.3568$, and $h_2^{(0)}=0.5279$. The inclusion of RC yields $h_0=0.0958$, $h_1=0.3532$, and $h_2=0.5209$ from the TBR and $h_{0F}=0.0005$, $h_{1F}=0.0031$, and $h_{2F}=0.0062$ from the FBR. With the value of $\tilde{\lambda}_+$ quoted above, we find that RC from the TBR induce a decrease of $0.8\%$ in the decay rate and the combined TBR and FBR effect produces an overall decrease of $0.26\%$, which is an important effect.

In summary, we have performed a detailed analysis of the computation of RC from the FBR of the Dalitz plot to order $(\alpha/\pi)(q/M_1)$, where $q$ is the momentum transfer and $M_1$ is the mass of the kaon. At this order of approximation, the expression obtained is model independent; it has been organized in such a way that its use in a Monte Carlo simulation is straightforward. The effects of the FBR are quite important when no discrimination of real photons can be made either kinematically or by direct detection; they produce a perceptible decrease in the total decay rate. Our results may constitute an important theoretical input in the determination of $V_{us}$ from $K_{l3}$ decays.

The authors acknowledge financial support from CONACYT and COFAA-IPN (M\'exico).

\end{document}